\documentclass[%
 amsmath,amssymb,
 aps,
]{revtex4-2}

\def\icarus{Icarus}

\def\aj{Astron.\ J. }
\def\apj{Astrophys.\ J. }
\def\apjl{Astrophys.\ J. }

\def\bssA_{Bull.\ Seismol.\ Soc.\ Am. }

\def\gcA_{Geochim.\ Cosmochim.\ Acta }

\def\mnras{Mon.\ Not.\ R.\ Astron.\ Soc. }

\def\nat{Nature }

\def\planss{Plan.\ Space Sci. }
\def\prA_{Phys.\ Rev.\ A }

\usepackage{graphicx}
\usepackage{dcolumn}
\usepackage{bm}

\begin{document}

\title{On the interstellar origin of high-inclination Centaurs}

\author{Fathi Namouni}
\email{namouni@oca.eu}
\affiliation{Universit\'e C\^ote d'Azur, CNRS, Observatoire de la C\^ote d'Azur, CS 34229, 06304 Nice, France
}

\author{Maria Helena Moreira Morais}
\email{helena.morais@rc.unesp.br}
\affiliation{Instituto de Geoci\^encias e Ci\^encias Exatas, Universidade Estadual Paulista (UNESP), Av. 24-A, 1515 13506-900 Rio Claro, SP, Brazil}

\begin{abstract}
Following our identification of the probable interstellar origin of high-inclination Centaurs, Morbidelli et al. \citep{Morbidellietal20} issued a rebuttal criticizing our methods and conclusions. Here, we show that the criticism is unfounded. Entropy's increase in the past is not an obstacle to accessing the statistical properties of Centaur past orbits as entropic expansion occurs around a time-independent conserved quantity that explains the probable orbits' clustering in parameter space, known as the polar corridor. The Copernican principle does not imply that  unstable motion does not exist in the solar system. It clarifies the meaning of the dynamical lifetime ensuring that Centaurs originating in the planetesimal disc are able to return to it in the time-backward simulation.  Our conclusions are supported by published independent research that shows conventional disc relaxation models do not explain the abundance of high inclination Centaurs and that enrichment from the interstellar medium is required. Future physical observations of the identified Centaurs and TNOs are likely to reveal the similarities and differences between solar system-born and interstellar origin Centaurs.
\end{abstract}

\maketitle


\section{Introduction}
Conventional solar system formation theory ascertains that the planets formed from a protoplanetary disc of gas and dust orbiting the Sun. When the gas dissipated, a thin planetesimal disc coplanar with the planets remained and extended no farther than Neptune's current orbit \citep{Pfalzner15}. This disc is thought to be the progenitor of Centaurs and transneptunian objects (TNOs) in the solar system \citep{Levison97,TiscarenoMalhotra03,Emelyanenko05,Disisto07,Brasser12,Nesvorny17,Fernandez18,Kaib19}.

Conventional solar system formation theory also ascertains that the solar system did not form in isolation from its Galactic neighbourhood. In the Sun's birth cluster, the various star systems with their planets and discs interacted and exchanged planetesimals and possibly planets with one another. It is thought that a significant component of the Oort cloud is likely of interstellar origin  \citep{Fernandez00,Levison10,Brasser12b,Jilkova16,Hands19,Kaib19}. 

The idea that some Sun-bound asteroids must be of interstellar origin is therefore not unconventional. It was only a matter of time before physical or dynamical evidence was found to recognize the interstellar origin of such objects. 

We provided such dynamical evidence for specific multiple-opposition high-inclination Centaurs in two articles \citep{NamouniMorais18b,NamouniMorais20b} hereafter Papers I and II respectively. Whereas our methodology requires the use of intensive computing, it is simple in principle. We take the orbit of a multiple-opposition high-inclination Centaur, clone it about a million times to cover the known observational uncertainty, then follow the evolution of the constructed clone swarm 4.5\,Gyr in the past.  As the method is statistical in principle, three outcomes are possible: first, the clones on average are scattered all over parameter space indicating that the method is inconclusive for the studied asteroid. Secondly, the clones on average cluster around the planets' common plane indicating that the asteroid probably originates from the planetesimal disc. Thirdly, the clones on average show clustering in parameter space indicating another probable origin. In Paper I we found that the majority of the 4.5\,Gyr clones of Jupiter's retrograde coorbital companion, asteroid Ka`epaoka`\=awela, clustered around a single orbit  inside Jupiter's Kozai-Lidov resonance in close proximity to  its current highly-inclined coorbital state. In Paper II, we found that 4.5\, Gyr-stable  clones of 19 high-inclination  centaurs  and TNOs cluster around a plane inclined by $90^\circ$ with respect to the solar system's invariable plane far outside Neptune's orbit.  The end states found in Papers I and II cannot originate from the planetesimal disc at $-4.5$\,Gyr as it was flat, confined inside Neptune's orbit and all its components orbited the Sun in the same direction as the planets. 

Recently Morbidelli et al. \cite{Morbidellietal20}  (hereafter called MBBR) claimed that our methodology is flawed and hence there was still no evidence for Sun-bound asteroids of interstellar origin. To support such claims, the authors raised three objections: the first is that the increasing entropy of a Centaur clone swarm in a time-backward simulation precludes access to any type of information, even statistically,  about the Centaur's original orbit. The second objection is related to the Copernican principle invoked in Papers I and II, which MBBR claim leads to a solar system without transient motion. The third objection is more personal belief than objection whereby MBBR  claim that high-inclination Centaurs must originate from the outer regions of the solar system which itself was populated by the early planetesimal disc. In essence, the interstellar origin is impossible. 

In this Letter, we examine these objections and explain why they are unfounded. In Section II, we show that entropy increase in the past is not an obstacle to accessing statistical information about original orbits because entropic expansion occurs around conserved quantities that explain the observed end-states. In Section III, we show that the Copernican principle does not remove the possibility of unstable motion in the solar system. It is crucial in ensuring that Centaurs that originated in the planetesimal disc can return to it in the time-backward simulation. In Section IV, we examine MBBR's personal belief that high-inclination Centaur originate from the planetesimal disc and how it is contradicted not only by our 20 million particle simulation but also by independent disc relaxation models MBBR chose to ignore. Section V contains our concluding remarks.

{Before we examine MBBR's objections, we point out that MBBR do not question the validity of our simulations and their setup. In effect, our work uses the same simulation methods that are commonly applied to study the dynamics and evolution of small outer solar system objects. With 20 million particles, our simulation in Paper II has achieved the highest resolution ever applied to the dynamics of outer solar system bodies.}

\section{The entropy objection}
{MBBR describe the Centaur clone swarm in our simulation by the tools of statistical mechanics because Centaurs reside in a chaotic region of phase space.} They say the accumulation of computational error increases the {\it swarm entropy  in the simulation} whether its motion is followed forward or backward in time. Next they consider  the {\it real} physical system not the simulated one whose entropy increase comes from  numerical error accumulation. This {real  clone swarm must obey the second law of thermodynamics. MBBR say its {\it real entropy} increases as time flows forward and decreases as time flows backward. {Disagreement between the two entropies of the real and the simulated swarms as time flows backward, 
MBBR say, means that there is no hope of accessing any information on a Centaur's original orbit not even statistically.} This conclusion would imply that the end states of the simulations in Papers I and II are not physical.

{To address this objection, we first explain how the real entropy of a Centaur clone swarm does not simply increase forward in time and nor  does the swarm's entropy simply increase in the time-backward simulation.}  Entropy measures disorder in a Centaur clone swarm, much as it does that of  a molecular gas  \citep{Feynman63}. It obeys the second law of thermodynamics that stipulates entropy of an isolated system should only increase. {However, a Centaur clone swarm   is not an isolated system and its entropy does not only increase.}  As the clones experience close encounters with the planets, their orbits are excited so that most collide with the Sun and the planets  or are ejected to the interstellar medium  whether we consider a real system or a simulated one. After an initial increase from excitation, entropy decreases drastically since  part of it is transferred to the collided objects as heat whereas another part is transferred to the interstellar medium as kinetic energy. {This occurs forward in time for a real swarm and both forward as well as  backward in time for a simulated swarm because the clones reside in a strongly chaotic region. If the number of clones is small (say a few hundred), the Centaur clone swarm's entropy in time reaches zero as all objects are removed from the swarm and disappeared from the solar system.}  In Papers I and II,  4.5\,Gyr long-lived orbits of high inclination Centaurs and TNOs were identified using large clone statistics. Within a Centaur clone swarm, the statistical ensemble composed of long-lived clones has an increasing entropy in a time-backward simulation. This increase is evidenced by the ensemble's expansion from the minuscule neighbourhood of the nominal orbit to the parameter space extents shown  in Fig. 1 and  2 of Paper I, Fig. 2 and 3 of Paper II and  Fig. 1 in this Section. {Summing up what occurs to a large clone swarm (with a million members), entropy in a time-backward simulation  starts  increasing as the clones are excited by the planets then decreases drastically on the dynamical timescale (1 to 100 Myr) \citep{TiscarenoMalhotra03,Horner04,Disisto07,BaileyMalhotra09,VolkMalhotra13} on account of the significant losses to collisions and ejections, then as only long-lived Sun-bound orbits remain, the swarm entropy tends to increase again.}

Next, we show that the logic of MBBR's objection is flawed as it leads to an absurd statement about the solar system. We do so using a simpler system than a Centaur clone swarm, whose origin is firmly rooted in the solar system and whose entropy evolution, when simulated, {invariably} increases forward and backward in time more in line with MBBR's logic.

We consider a million clone swarm of the Earth in a minuscule neighbourhood of its known orbit and simulate its evolution back in time  to $-3$\,Gyr.\footnote{We choose 3\,Gyr to stay well clear of the epoch when the planet's orbit could be influenced by planet migration.} 
It is   well established that the motion of the inner solar system planets is chaotic \citep{Laskar89,Laskar90} and that their orbits experience chaotic  diffusion in phase space \citep{Laskar08}. The Earth clone swarm can therefore be treated statistically. As the Earth's orbit is secularly stable, the simulated swarm's entropy increases forward and backward in time as there are no collisions or ejections in  this case in agreement with  the setting of MBBR's objection. Then following MBBR's logic we should compare such a simulated swarm to a real Earth clone swarm that must obey the second law of thermodynamics so that its entropy increases in the future and decreases in the past. Since the simulated and real entropy changes disagree when time flows backward, MBBR tell us all information about Earth's orbit at  $-3$\,Gyr is lost implying that whatever is seen at the simulation's end is physically meaningless. 

Without resorting to complex mathematical analyses, common sense indicates that Earth clones, $3$\,Gyr ago, will still be located near the planet's  present orbit despite numerical error accumulation and that the statistical properties of the clones inform us about the Earth's mean orbital elements as well as their chaotic diffusion. {This is evidently confirmed by the 5\,Gyr time-backward simulation of the solar system's  planets by Laskar \cite{Laskar08} using 1001 solar system clones (see also \cite{Ito02}).}

The absurd conclusion of MBBR's logic about the Earth clone swarm could have been avoided if they had excluded the mention to {\it any information even statistically} from their statement, {and had only sustained that the disagreement between the real and the simulated entropies precluded access to the original orbit.} However that would not have made their case since at no point in Papers I and II  do we claim to determine a Centaur's original orbit {as  only   the statistical distributions of Centaur end states  are derived at $-4.5$\,Gyr (see eg. Fig. 3 of Paper II).} Not even the most sophisticated solar system ephemeris can achieve the feat of determining an original orbit on Gyr timescales. Only statistical methods, such as those in \cite{Laskar08} and Papers I and II, can be used to ascertain the long term past dynamics of solar system objects.

The example of the million Earth clone swarm is also useful to understand why the Centaur end states we find are physically meaningful. Earth's orbit is known to be stable.  Its long term dynamics are modeled by the secular part of the planetary disturbing function \citep{Laskar89}. This is possible as the Earth's semi-major axis is conserved. This mean that the Earth clones will cluster around the present semi-major axis as it is a conserved quantity. The Centaur end states in Papers I and II do not scatter all over parameter space {as would MBBR's objection have required with the loss of all information}. Instead, they exhibit clusterings as explained in the Introduction. Are these clusterings related to conserved quantities? The answer is affirmative. 

First encountered in Paper I (see Fig. 1 therein), the Tisserand relation applied to Ka`epaoka`\=awela's was shown to describe accurately the mean path that the clone distribution follows in the inclination-semimajor axis plane.  The region of physical space associated with that curve was termed `the polar corridor' and its identification was the reason for our interest in the origin of high-inclination Centaurs in Paper II. The effect of the conserved quantity was first seen but not recognized in the discovery paper of TNO  (471325) 2011 KT19 where the stability of its orbit was simulated using 1000 clones over 1\,Gyr \citep{Chen16}.  The conserved quantity in the inclination-semimajor axis plane is nearly identical to the Tisserand relation only when one major perturber  influences the motion of the clone swarm. This is understandable since the Tisserand relation originates from the conservation of the Jacobi constant of the three-body problem  \citep{ssdbook}  whereas our simulation has four perturbing planets. {We give three new examples of Centaurs studied in Paper II that illustrate this concept.} Centaur 2016 YB13 is currently in Jupiter's coorbital region with a 140$^\circ$ inclination and an eccentricity of 0.41.  Its major pertuber is Jupiter and its clone distribution at $-10$\,Myr is centred around the Jupiter-Tisserand relation curve shown in Fig. 1 (left panel). Similarly, TNO (471325) 2011 KT19 is on a nearly-polar orbit with 110$^\circ$ inclination in the 7:9 resonance \citep{MoraisNamouni17b} outside Neptune's orbit whose Tisserand relation the clone swarm follows as shown at  $-1.2$\,Gyr in Fig. 1 (right panel). Retrograde Centaur 1999 LE31 is located between Jupiter and Saturn with a semimajor axis of 8.13\,au, a $150^\circ$   inclination and an eccentricity of 0.47. The clone distribution shown in Fig. 1 (middle panel) at $-50$\, Myr has  a similar inclination-semimajor axis  structure as those of the previous two objects but the average curve is not given by Jupiter's nor Saturn's Tisserand relations with the Centaur. {In all cases, the Tisserand-like conserved quantity confines the clones' inclination to 90$^\circ$ with a small inclination dispersion and a large semi-major axis one extending from the scattered disc to the Oort cloud regions and explaining Centaur end states and their probable interstellar origin. Recently, K\"ohne and Batygin \cite{kohne20} simulated 6200 clones of Ka`epaoka`\=awela 100\,Myr in the past and confirmed the clone clustering in the polar corridor around the Tisserand relation.} 

Centaur Ka`epaoka`\=awela's clones that cluster in Jupiter's 1:1 resonance are also associated with conserved quantities. In addition to clustering around Jupiter's semi-major axis, the clones are all locked in the Kozai-Lidov resonance with a small libration amplitude, implying that both the vertical components of their angular momenta as well as their arguments of perihelion are conserved (Paper I).  

{The conserved quantities of Tisserand and Kozai-Lidov have another important implication for the clone clustering observed in our simulations. Since they are time-independent, the conserved quantities confine the evolution of the long-lived clones in the inclination-semimajor axis plane  whether evolution is followed backward or forward in time. This explains why the end states clustered in the polar corridor will still evolve inside it in a time-forward simulation (see Section IV for another consequence of this property.) }

\begin{figure*}
\hspace*{-10mm}\includegraphics[width=175mm]{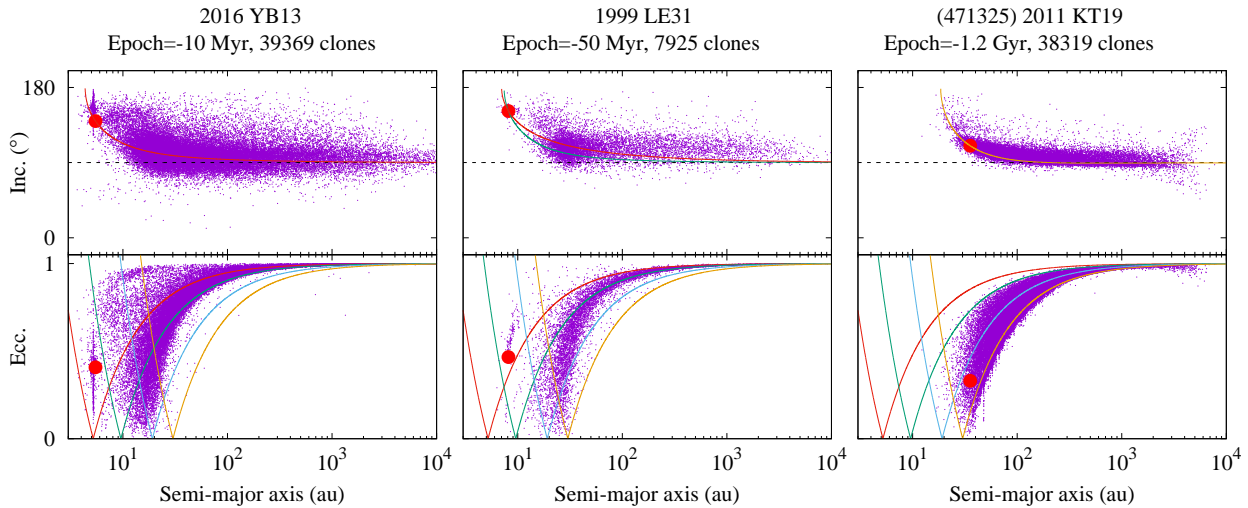}\\[-25mm]
\caption{\label{fig1}Clone distributions in the semi-major axis-inclination and  semi-major axis-eccentricity planes of  2016 YB13 (left panel), 1999 LE13 (middle panel) and (471325) 2011 KT19 (right panel).  The initial clone number is $10^6$ for the first two Centaurs and $1.2\times 10^5$ for the TNO. The filled red circles indicate the location of the nominal orbit at the current epoch. The snapshot epoch and the number of clones present are indicated above each panel. The curves in the inclination-semi-major axis plots correspond to 2016 YB13's Tisserand relation with Jupiter (left panel), 1999 LE13's Tisserand relation with Jupiter (top curve, middle panel) and that with Saturn (bottom curve, middle panel) and  TNO (471325)'s Tisserand relation with Neptune. The four V-shaped curves in the eccentricity-semimajor axis plots  denote the intersection at aphelion and perihelion of the clone's orbit with those of Jupiter, Saturn, Uranus and Neptune respectively from left to right.}
\end{figure*}

\section{The Copernican principle objection}

{MBBR claim that our use of the Copernican principle in Papers I and II implies that transient or unstable motion becomes impossible in the solar system. In the following, we explain why the Copernican principle is needed and why unstable motion is always possible. }

To the question: how old are the solar system's Centaurs? MBBR's answer will be 4.5\,Gyr. We know so because in their third objection, they argue for a  planetesimal disc origin for all Centaurs (see Section IV).  {But in Papers I and II, we also reply 4.5\, Gyr to the same question.} Does MBBR's answer imply that unstable motion does not exist in the solar system? Clearly, no. Similarly, our answer does not imply unstable motion is inexistent. 

If all simply agree that Centaurs were present in the solar system for 4.5\,Gyr whether they originated from the interstellar medium or the planetesimal disc, why is there a need to  invoke the Copernican principle? The reason is that the simulation, by reflecting the complexity of Centaur dynamics, offers strong enticements to recognize the unstable orbits as the true past orbits of all Centaurs of low and high inclination alike. If unstable orbits in time-backward simulations do represent Centaur evolution then the possibility of originating from the planetesimal disc no longer exists and all Centaurs come exclusively from the interstellar medium. The Copernican principle, applied to the meaning of the dynamical lifetime, ensures that the planetesimal disc origin remains possible and  that Centaurs originating from the planetesimal disc are able to return to it at the end of the 4.5\,Gyr time-backward simulation.

{To understand this, we recall that when a Centaur clone swarm is simulated forward or backward in time, two types of orbits are found.} Short-lived orbits whose dynamical  lifetimes range from 1 to 100 Myr. These are the unstable orbits. The second type is long-lived orbits that reach the end of planet formation and are therefore 4.5\,Gyr old. These are the stable orbits.

A Centaur's orbit is known only with finite accuracy. In order to cover its observational uncertainty and sample a mere 10 points within the error bar of a single orbital element, a million clone orbits are needed as phase space has six dimensions (3 positions and 3 velocities).  The 4.5\, Gyr long-lived planetesimal disc Centaurs {\it must} have crossed the strongly chaotic sea between the giant planets and occupied its stable islands on their way from the  planetesimal disc to their current positions otherwise they would {\it never} have survived to the present. Such islands are quite hard to find so much so that the time-backward simulation {offers a choice between} a few tens of  4.5\,Gyr long-lived orbits against nearly a million short-lived ones, the number disparity reflecting how rough it was to navigate the chaotic sea with an initially uncertain Centaur orbit. 

{The short-lived orbits do not stop at numbers and present an additional enticement in the form of  a clear unified origin for all Centaurs regardless of inclination.} As Sun and planet collisions are unphysical in time-backward simulations, ejected orbits are the only possible short-lived orbits. Since ejection from the solar system in a time-backward integration means capture from the interstellar medium when the arrow of time is restored to its natural direction, nearly half a million\footnote{The other half are the discarded unphysical ones.} short-lived clones in the simulation are literally seen  joining the solar system  from the interstellar medium\footnote{The unstable orbits of high inclination Centaurs join the solar system through the polar corridor. Low inclination Centaurs do not as the example of the fictitious Chiron shows in Paper II whose clones join the solar system $30^\circ$  above the invariable pane.}.  Meanwhile, there are always a few tens of 4.5\, Gyr long-lived orbits that successfully navigated the chaotic sea and claim to represent the original Centaur whether it came from the planetesimal disc or the interstellar medium.

The half million unstable clone orbits therefore make the strongest possible case for the interstellar origin of all Centaurs, high and low inclination alike, because regardless of inclination, Centaur past dynamical lifetimes range from 1 to 100\,Myr and never hope to reach 4.5\,Gyr.  The planetesimal disc is no longer an origin option for any Centaur. 

Ruling out the planetesimal disc as a possible origin for solar system Centaurs is not acceptable. This means some element  about the short-lived unstable orbits was taken for granted or is missing in the previous analysis. That element is the meaning we assign to the past dynamical lifetime as the true past lifetime of a Centaur. As the $N$-body gravitational problem is time-reversible, the past and the future dynamical lifetimes of the short-lived orbits of a given Centaur are nearly identical --we denote their value by  $T_d$(Centaur) \citep{Horner04,Port18}. This means that every Centaur with its specific $T_d$ must have joined the solar system from the interstellar medium near epoch $t=-T_d$(Centaur) with respect to the present and will suffer an ejection or a collision with a solar system body at epoch $t=T_d$(Centaur). This scenario gives the present epoch $t=0$ a special significance in solar system history as the capture process and the onset of ejection or collision of each Centaur should be fine tuned in order for the Centaur's  $2T_d$(Centaur) lifetime in the solar system to be centered on the present. That is not allowed by the Copernican principle. 

The Copernican principle thus implies that the true past (and future) lifetime of a Centaur must be larger than the dynamical lifetime. The dynamical lifetime is actually a measure of how rough navigating the chaotic sea can be and so does the disproportionately large number of short-lived orbits with respect to that of 4.5\,Gyr orbits. As the true past lifetime of a Centaur can be arbitrarily larger than the dynamical lifetime, it may even equal the solar system's age thereby leaving the door open for 4.5\, Gyr long-lived Centaurs to originate from the planetesimal disc. Similarly, the true future lifetime is larger than the dynamical lifetime and can accommodate the possibility of an ejection or a collision. At no time in this analysis does the Copernican principle force us to assume or leads to a solar system with no transient or unstable motion.

\section{The interstellar origin objection}
{MBBR claim that our work did not account for the possibility that high inclination Centaurs may have originated from the Oort cloud and that the interstellar origin is impossible. MBBR argue that the dynamics of such Centaurs resemble those of Oort cloud comets but they admit they cannot account for their smaller semi-major axes. 

It should first be noted that at $-4.5$\,Gyr the Oort cloud was empty of solar system material but must have captured a sizable fraction of interstellar objects from the Sun's birth cluster (see Section I). When MBBR mention an Oort cloud origin, they actually consider the conventional view that suggests that, at the end of planet formation $4.5$\, Gyr ago, the planets scattered planetesimals from the disc to the Oort cloud region in the first Gyr or so, and that in time, such objects return from the Oort cloud and assume Centaur orbits among the giant planets. MBBR's so-called `Oort cloud origin' is actually a planetesimal disc origin when the origin epoch is set to $-4.5$\,Gyr. This is why MBBR's `Oort cloud origin' excludes the possibility that high inclination Centaurs can be of interstellar origin. 

Two independent research works  examined whether the abundance of high inclination Centaurs can be explained by the conventional models of the planetesimal disc's relaxation from $-4.5$\, Gyr to the present and answered negatively \citep{Kaib19,Nesvorny19}. Both works are mentioned in Paper II. In their Letter, MBBR ignore the findings of Kaib et al. and claim that the analysis of Nesvorn\'y et al. lacks resolution. Kaib et al. concluded that the abundance of high inclination Centaurs cannot be explained by the planetesimal disc with or without a hypothetical outer planet and that enrichment from the interstellar medium can resolve this problem.

Furthermore, in Papers I and II, the statistical distributions of Centaur end-states  extend from the scattered disc to the inner Oort cloud indicating that their original radial location may be in the Oort cloud at $-4.5\,$Gyr. As the Oort cloud is believed to have been empty of solar system material and must have sheltered a sizable planetesimal population captured from the interstellar medium, our results are compatible with an early interstellar planetesimal  capture occurring in the inner Oort cloud region.  

Papers I and II  can even solve the obstacle MBBR encounter when trying to argue that high inclination Centaurs originate from the Oort cloud but that their semi-major axes are inconveniently smaller than those of Oort cloud comets. Our 20 million particle simulation shows that the solution to MBBR's problem is the polar corridor. In a time-forward simulation, Centaurs arriving from the Oort cloud (whatever their arrival epoch) that end up with high inclinations enter safely the giant planets domain if they have nearly polar orbits (see Section II).   

MBBR claim that our methodology does not consider different outcomes other than an interstellar origin as though we had fine-tuned a simulation to support exclusively a specific outcome. In actuality, our method takes the Centaur clone swarm constructed in agreement with the known observational errors and let it evolve freely for 4.5\,Gyr. The simulation has only one parameter, the Centaur clone number. If that number is sufficiently large, as in Papers I and II, the outcome, namely the statistical distributions of Centaur end states, is unique. There is no possible way to fine-tune the simulation and aim it towards a different outcome in contrast to solar system formation models whose initial state is postulated and parametrized by a number of variables that are subject to revision (see e.g. \cite{Ribeiro20}). }

\section{Conclusion}
Upon close examination, the so-called `fatal flaws' seen by MBBR in our work vanish. {Entropy is not an obstacle to accessing statistical information about Centaurs' original orbits as clone evolution is confined by time-independent conserved quantities such as the Tisserand relation or the Kozai-Lidov constant. The Copernican principle does not advocate a solar system with no unstable motion. It clarifies the meaning of the dynamical lifetime ensuring that planetesimal disc Centaurs are able to retrace their steps in the past. MBBR's critique is motivated by their belief in a planetesimal disc origin for high-inclination Centaurs. Such an origin has been disproved by two independent peer-reviewed works based on the relaxation of the planetesimal disc.}   There is however is no reason why other solar system Centaurs cannot originate from the planetesimal disc  \citep{Kaib19,Nesvorny19}. It is, in fact, our hope that physical studies of the objects we identified will shed some light on the similarities and differences with low inclination Centaurs \citep{Peixinho19} in order to be able to physically distinguish  solar-system born from interstellar origin Centaurs.

\section*{Acknowledgments}
The simulations shown in Fig. 1 were done at the M\'esocentre SIGAMM hosted at the Observatoire de la C\^ote d'Azur. 
M.H.M. Morais research had financial support from S\~ao Paulo Research Foundation (FAPESP/2018/08620-1) and CNPq-Brazil (Pq2/304037/2018-4). This research was supported in part by FINEP and FAPESP through the computational resources provided by the Center for Scientific Computing (NCC/GridUNESP) of the S\~ao Paulo State University (UNESP).

\bibliography{0ms}

\end{document}